\newif\ifpreprint                                       
\preprintfalse                                          

\ifpreprint                                             
 \documentstyle[prl,aps,epsfig,preprint,12pt]{revtex}
\else                                                   
 \documentstyle[prl,aps,epsfig,twocolumn]{revtex}       
\fi                                                     

\newcommand{\G}{\Gamma}
\newcommand{\w}{\omega}
\newcommand{\g}{\gamma}
\newcommand{\s}{\sigma}

\newcommand{\E}{\mbox{${\cal E}$}}

\newcommand{\Gt}{\tilde{\Gamma}}

\newcommand{\me}{|e_1|}

\newlength{\textwidthm}
\begin{document}
\draft

\ifpreprint                                             
 \preprint{Inversionless IR generation}
\else                                                   
 \wideabs{                                              
\fi                                                     

\title{
Novel infrared generation in low-dimensional \\
semiconductor heterostructures via quantum coherence
} \author{A. A. Belyanin$^{1,2}$, F. Capasso$^3$, 
V.  V.  Kocharovsky$^{1,2}$, Vl. V. Kocharovsky$^2$,
and M.O. Scully$^{1,4}$
}
\address{$^1$Physics Department and Institute for Quantum 
Studies, \\
Texas A\&M University, College Station, TX 77843
 \\
$^2$ Institute of
Applied Physics , Russian Academy of Science, \\ 46 Ulyanov 
Street,
603600 Nizhny Novgorod, Russia \\
$^3$Bell Laboratories, Lucent Technologies, \\ 600 Mountain 
Avenue,
Murray Hill, NJ07974, USA \\
$^4$Max-Planck Institute Fur Quantenoptik, 85748 Garching, 
Germany
}

\date{\today}
\maketitle
\begin{abstract}
A new scheme for infrared  generation without
population inversion between 
subbands in quantum-well and quantum-dot lasers is presented 
and documented by detailed calculations. 
The scheme is based on the simultaneous 
generation at three frequencies:
optical lasing at the two  interband
transitions which take place simultaneously, in the same active 
region, and serve as the coherent drive for the IR field. 
This mechanism for frequency
down-conversion does not rely upon any ad hoc assumptions of  
long-lived coherences in the semiconductor active medium. And it should 
work efficiently
 at room temperature with injection current pumping. 
For optimized waveguide and cavity parameters, 
the intrinsic efficiency of 
the 
down-conversion process can reach the limiting quantum value 
corresponding to one infrared photon per one optical photon. 
Due to the parametric nature of IR generation, the proposed inversionless 
scheme is especially promising for long-wavelength (far-
infrared) operation.

\end{abstract}

\ifpreprint                                             
\bigskip\bigskip                                        
\fi                                                     

\ifpreprint                                             
 \newpage
\else                                                   
 } 
-
 \narrowtext                                            
\fi                                                     

Low-dimensional semiconductor heterostructures are almost 
ideally suited for
generation in the mid- to far-infrared range (denoted below as IR for
brevity), because the spacing  between levels of dimensional 
quantization
can be conveniently manipulated  over the region from several to
hundreds of microns, and injection pumping 
is possible. There is however a major problem:  
strong non-resonant losses of the IR field due to 
free-carrier absorption and diffraction, which become increasingly 
important at longer far-IR wavelengths. Due to very short lifetime of 
excited states, it is difficult to maintain a large enough population 
inversion and high gain at the intersubband transitions, necessary 
to overcome losses. 
There were many suggestions to solve this
problem by
rapid depletion of the lower lasing state using, e.g., the resonant
tunneling to adjacent semiconductor layers or transition to yet 
lower subbands due to phonon emission \cite{KS71,F94},
or even stimulated interband recombination \cite{S96}; see 
\cite{K99,L98} for recent reviews. The successful culmination of 
these studies is the realization of quantum cascade lasers \cite{F94}, 
in which the lower lasing state is depopulated either by tunneling in 
the superlattice or due to transition to lower-lying levels separated 
from the lasing state by nearly the energy of a LO-phonon; see e.g. 
\cite{C99} and references therein.

We here put forward another possibility, allowing us to achieve IR
generation  without population inversion at the intraband transition. 
This
becomes possible with the aid of  laser fields simultaneously  
generated at
the
{\it interband} transitions (called optical fields for brevity),
which  serve as the coherent drive for the frequency down-
conversion to
the IR. Employing self-generated optical lasing fields provides the 
possibility of injection current pumping and also removes the 
problems associated
with external drive (beam overlap, drive absorption, spatial 
inhomogeneity), which were inherent in previous works on 
parametric down-conversion in semiconductors, see \cite{K99} for a 
review. 
The second important feature is the great enhancement of nonlinear 
wave mixing near resonance with intersubband transitions. 
The third  feature is the possibility of  canceling the resonance one-photon 
absorption for the generated IR field due to coherence effects provided 
by self-generated driving optical fields \cite{lwi}.  The processes 
incorporating these three features constitute an important field of 
research with a variety of physical effects and promising 
applications. 
We consider just one example of such processes of IR generation 
and nonlinear mixing with self-generated optical fields in 
semiconductor heterostructures. 

To avoid misunderstanding, we note that the mechanism of IR generation 
discussed in this paper is different from the approach in
which the resonant tunneling and Fano-type interference are used 
to establish a large coherence at intersubband transitions 
\cite{SN99}. These 
quantum interference ideas  imply usually the presence of a long-
lived coherence at the intraband
transitions. Our approach does not require long dephasing times. 
We here focus on nonlinear wave mixing phenomenon,
which is greatly enhanced near the resonance with intersubband 
transition. 
Note that this process has some common features with recently  
observed generation of coherent IR emission in rubidium vapor in 
four-wave mixing experiments \cite{Z00}. Also,  coherent microwave 
generation at the difference frequency
under the action of two resonant  external optical fields was 
observed in cesium vapor \cite{VG98}.

{\bf Generic three-level scheme.}
As the simplest case, consider the situation when only three levels 
of dimensional quantization are involved in generation: one 
(lowest-lying) heavy-hole level, and two electron levels; see Fig.~1. 
Of course, this scheme also describes the situation when there 
are two hole levels and one electron level involved. We need all 
three transitions to be allowed by selection rules. 
In a quantum well (QW) this will 
generally require using asymmetric structures, e.g. rectangular well 
with different barrier heights. 
For example, in a 
Al$_{0.3}$Ga$_{0.7}$As/GaAs/Al$_{0.2}$Ga$_{0.8}$As QW 
parametric IR 
generation is possible either between two electron subbands $1e$ 
and 
$2e$ separated by 98 meV ($\lambda \simeq 13$ $\mu$m) or 
between two 
lowest heavy-hole subbands ($\lambda \simeq 60$ $\mu$m).  

Symmetric QWs can also be employed. 
e.g.  in the case of 
a strong coupling between different subbands of heavy and light 
holes. A typical example is 
Al$_{0.3}$Ga$_{0.7}$As/GaAs/Al$_{0.3}$Ga$_{0.7}$As 
QW in which the second subband of heavy holes $2hh$ and the 
first subband of light holes $1lh$ 
happen to be very close to each other 
(within homogeneous linewidth) in the $\Gamma$-point  for a wide 
range of thicknesses $\sim 5-8$ nm, 
and therefore are strongly mixed. 
In this case two optical fields correspond to $1e\rightarrow 1hh$ 
and $1e\rightarrow 1lh$ transitions. And IR field is generated 
via the $1hh \rightarrow 2hh$ transition.  

A third configuration of interest is the quantum dot (QD). 
For example in a self-assembled InAs/GaAs quantum dot (QD) the 
three-level scheme can be easily realized with all three transitions 
allowed \cite{B98}. 

When the injection current density reaches the threshold value  
$j_{\rm th}$, optical generation starts 
due to recombination transitions 
between ground electron and hole states. Upon  increasing the 
pumping current, 
optical generation can start also from excited states and  the 
laser can be completely switched to lasing from 
the excited-state which 
has higher maximum gain due to a larger density of states.  The effect 
of 
 excited-state lasing was studied both in QW and QD lasers 
 \cite{K99,B98,T86,C92}. It was found that with optimized laser 
 parameters the region of simultaneous ground-state and excited-
state 
lasing can be around $j \sim 2 j_{\rm th}$ \cite{T86,C92}. In order 
to have the region of two-wavelength lasing sufficiently broad 
($\Delta j \sim (0.1-0.2)  j_{\rm th}$), gains for the two 
wavelengths should be close to each other. 

The presence of one or two strong optical driving fields in the cavity
gives rise to a rich variety of {\it resonant} coupling mechanisms
by which the IR field can be produced.
Here  we will concentrate on one such scheme
in which the two coherent optical fields having frequencies $\w_1$ 
and
$\w_2$ excite the induce electronic oscillation
at the difference frequency $\w_2-\w_1$.
It is important to note that 
the coherent IR polarization is parametrically excited
independent of the sign of population difference 
at the IR transition.

The resulting output intensity of IR radiation depends on the 
coupling coefficient between the IR polarization and the cavity modes. 
It 
 is clear that  polarization wave has longitudinal wavenumber $k_x$ 
equal to the difference $k_{2x}-k_{1x}$ of longitudinal wavenumbers 
of the two optical fields. Therefore, only the mode having the above 
wavenumber is efficiently excited. The field intensity is maximized 
 when the frequency of this mode is equal to the difference 
frequency 
 of optical fields. This requires special waveguide design since 
 refractive indices of bulk semiconductor materials for optical and 
 IR frequencies are different.  For far-IR generation, there is more 
flexibility due to efficient manipulation of the refractive index by 
a slight doping. 

{\bf Basic model}.
To quantify the above ideas, 
we have calculated the excited IR polarization 
and field by solving the coupled electronic 
density-matrix equations and electromagnetic Maxwell 
equations for the three fields, assuming steady-state.  
It is convenient to expand all fields in 
an orthonormal set of cavity modes ${\bf F}_{\lambda}$ 
and to introduce slowly varying complex amplitudes of fields and 
 polarizations. For example, for the IR field at the $3\rightarrow 2$ 
 transition we can write \begin{equation} \label{1} {\bf E}({\bf 
r},t) = \sum_{\lambda} \frac{1}{2} {\E}(t) {\bf F}_{\lambda}({\bf r}) 
\exp(-i\w t) + {\rm c.c.} \end{equation} We will assume that the 
mode 
has a simple $\exp(\pm ik_x x)$ dependence in the propagation 
direction $x$ with  the refractive index $\mu = k_xc/\omega$ and 
the 
 transverse structure defined by a  specific waveguide. 

After introducing complex
Rabi frequency $ e(t) = d \E(t)/2\hbar$, the wave equation
can be written as
\begin{equation} \label{2}
 \frac{de}{dt} + (\kappa + i(\w_c-\w))e =  \frac{2\pi i \w d^2 N}{\hbar
\mu^2}
\int_{V_c} \sum_j \s_{32}^j F_{\lambda}({\bf r})\; d{\bf r}.
\end{equation}
Here $d$ is the dipole moment of the IR transition, $N$ is the total
volume density of electron states in the active region, $\w_{32}$ the 
central frequency of
$3 \rightarrow 2$ transition, $\w_c$ the frequency of the IR cavity 
mode
with given $k_x$,
$\kappa$ the cavity losses, $V_c$ the cavity volume.  The variable  
$\s_{32}^j$ is the slowly varying amplitude of the element 
$\rho_{32}^j$ of density matrix, index $j$ labels different electron 
states contributing to the inhomogeneously broadened line. For a 
system of QDs, $j$ is simply the dot label. In QWs, index $j$ 
labels 
different ${\bf k}_{\parallel}$ states with respect to longitudinal 
quasimomentum. Only ${\bf k}_{\parallel}$-conserving transitions 
are 
considered, where ${\bf k}_{\parallel}$ lies in the plane of the 
wells ($xy$-plane).

The same representation is assumed for
the two optical fields ${\bf E}_{1,2}$, with off-diagonal density matrix 
elements $\s_{21}^j$, $\s_{31}^j$ on the right-hand side, and 
corresponding parameters $\mu$, $d$, $\w$, $\w_c$, $\kappa$, 
$k_x$, 
$g$  having index 1 or 2.  Note that the optical fields can have {\it 
arbitrary} polarization with respect to the parametrically excited IR 
field. In particular, for a QW laser the IR field should be 
$z$-polarized in the general case, while the optical fields are 
preferentially $y$-polarized. 

 Expressions for $\s_{ik}$  and population differences
$ n_{ik} = \rho_{ii} - \rho_{kk}$,  $i,k = 1,2,3$, 
 are found from the density matrix equations
 with phenomenological rates of relaxation and pumping.
For simplicity, we will assume bipolar injection with equal injection 
rates of electrons to level 3 and holes to level 1, so the total 
particle density is conserved.
In the limit $|e| < \g_{ik}$ the
 resulting amplitude of the IR field is found to be
\begin{equation} \label{6}
e \simeq \frac{i g^2 e_1^* e_2}{\kappa}
\sum_j \left(\frac{n_{12}(\nu_j)}{\G^*_{21}\Gt_{32}}
+ \frac{n_{13}(\nu_j)}{\G_{31}\Gt_{32}}\right),
\end{equation}
where
$ g^2 = 2\pi  \w d^2 N G/( \hbar \mu^2)$, 
\begin{eqnarray}
 \G_{21} &=& \g_{21} + i(\w_{21} + \nu_j - \w_1), \nonumber\\
\G_{31} &=& \g_{31} + i(\w_{31} + \nu_j - \w_2), \nonumber \\
\G_{32} &=& \g_{32} + i(\w_{32} + \nu_j - \w_2+\w_1), \nonumber \\
\Gt_{32} &=& \G_{32} + |e_1|^2/\G_{31} + |e_2|^2/\G^*_{21} 
\nonumber.
\end{eqnarray}
Here $G$ is the optical confinement factor for the IR field, $\nu_j$ 
is the difference between the transition frequency for a given $j$th 
state and the central frequency $\w_{21}$, $\w_{31}$, or $\w_{32}$. 
 
The resonance $\w_c = \w_2-\w_1$ with a given cavity mode having 
the 
wavenumber $k_x \simeq k_{2x}-k_{1x}$ is assumed.

The value of the dipole moment at the IR transition is typically $d 
\sim (1-3)$ nm, while it is $(0.3-1)$ nm for the optical  transitions 
\cite{K99,B98}.  The relaxation rates $\gamma_{ik}$ of both optical 
and IR polarizations are of order 5-10 meV  in QW lasers at room 
temperature (i.e., the relaxation time $\lesssim 0.1$ ps) 
\cite{CK99,A89,H99} and can be several times lower in self-
assembled 
QDs (1 ps).  In the IR range the cavity losses are mainly due to 
free 
 carrier absorption. At the high carrier densities $N \gtrsim 10^{18}$ 
 cm$^{-3}$ necessary for the excited-state lasing in QW's the 
material 
 losses in a bulk active medium are of order 100 cm$^{-1}$ at 
 $\lambda \simeq 6$ $\mu$m, and grow as $\lambda^2$ or 
$\lambda^3$ 
 depending on the dominant scattering mechanism \cite{J85}. Thick 
 cladding layers have smaller doping density of order 
 $4\times 10^{16}-10^{17}$ cm$^{-3}$, but can contribute 
 significantly to the losses due to the large  overlap factor 
 $G \sim 1$.  Large losses are a major problem in all 
 proposed far-IR lasing schemes involving free carriers in 
 semiconductors. Note, however, that increasing losses is not a 
 principal limitation in the present parametric scheme since they do 
 not prohibit generation itself, but rather decrease the  IR field 
intensity. This possibility of far-infrared operation is an important 
advantage of the proposed mechanism. Note also that in QD lasers 
the 
excited state lasing can be achieved at lower carrier densities due 
to the state filling effect, and the intrinsic losses in active 
medium are lower.

{\bf Homogeneous broadening.}
This case is relevant for QWs at low temperatures. For room 
temperature, the quasi-Fermi energy for electrons is expected to 
be 
larger than the homogeneous bandwidth of 7-10 meV, and 
inhomogeneous 
broadening cannot be neglected. As for the self-assembled QDs, 
present-day structures have a large inhomogeneous broadening 
$\gtrsim 
20$ meV associated with the spread of dot sizes, which is 
definitely 
much larger than the homogeneous linewidth.

The resulting expression for the IR field is 
\begin{equation} \label{7}
|e| \simeq \frac{|e_1||e_2|}{\g_{32}}
\left( \frac{\w}{\w_1} 
\frac{d^2}{d^2_1}\frac{\kappa_1}{G_1}\frac{G}{\kappa}
+  \frac{\w}{\w_2} 
\frac{d^2}{d^2_2}\frac{\kappa_2}{G_2}\frac{G}{\kappa}
\right),
\end{equation}
where $G$, $G_1$ and $G_2$ are the IR and optical confinement factors. 
At a given wavelength, the crucial parameter in Eq.~(\ref{7}), which 
governs the efficiency of down-conversion, is $\eta = 
 (\kappa_{1,2}/G_{1,2}) (G/\kappa)$. 
The main source of IR losses is free-carrier absorption. 
For the optical fields the ratio $\kappa_{1,2}/G_{1,2}$ 
in which the material gain at the optical transition, 
which is of order $10^3$ cm$^{-1}$ in QWs and $10^4$ cm$^{-
1}$ in QDs. 
Therefore, even  for very high IR material losses of order  
$10^3-10^4$ cm$^{-1}$, 
the $\eta$ parameter can still be close to 
unity. As we already mentioned, IR losses can be  
dominated by absorption in doped cladding layers. 
Let us take as an example $2\kappa \simeq 150$ cm$^{-1}$ as 
measured in 
quantum cascade lasers at 17 $\mu$m wavelength. If  
$2\kappa_{1}/G_{1} \simeq 1500$ cm$^{-1}$, then we have $\eta 
\sim 1$ 
for $G \sim  0.1$. Evidently, the value of $\eta$ decreases rapidly 
with increasing wavelength due to growing $\kappa$ and 
decreasing 
$G$.

{\bf Inhomogeneous broadening.}
 We will assume here that the inhomogeneous widths $u_{ik}$ of 
all transitions
are much larger than all homogeneous bandwidths $\g_{ik}$ and 
Rabi frequencies $|e_{1,2}|$  of optical 
laser fields that are at  exact resonance with the centers of 
inhomogeneously broadened lines: $\w_1 = \w_{21}$ and $\w_2 = 
\w_{31}$. We 
 consider two different situations  in which  explicit formulas can be 
obtained: when the optical field intensities are (i) much smaller and 
(ii) 
much greater than the saturation
values. 

{\bf (i)}. In this case the populations have the spectral 
distributions as supported by pumping in the absence of generated 
fields (no spectral hole burning).  When $u_{ik} \gg \g_{ik}$, the 
 precise shape of the inhomogeneous line is not important and  
simple 
 expression for the IR field can be obtained:  \begin{equation} 
\label{10a} |e| \simeq \frac{2 |e_1||e_2|}{(\g_{32}+ \g_{21})} 
\frac{u_{21}}{u_{32}} \frac{\w}{\w_1} 
\frac{d^2}{d^2_1}\frac{\kappa_1}{G_1}\frac{G}{\kappa}, 
\end{equation} 
or, in terms of intensities, \begin{equation} \label{11} |\E|^2 
 \simeq  |\E_1|^2 \frac{|\E_2|^2}{|\E_2|^2_s} \left( \frac{2 
\g_{32}}{(\g_{32}+\g_{21})} \frac{d}{d_1}\frac{\w}{\w_1} 
\frac{u_{21}}{u_{32}} \frac{\kappa_1}{G_1} \frac{G}{\kappa}\right)^2.  
\end{equation} Here $|\E_2|^2_s = \hbar^2 \g_{32}^2/ d_2^2$. 
Equation 
(\ref{10a}) is similar to (\ref{7}) except for the ratios of 
inhomogeneous linewidths that are expected to be of the order of 
unity. As we see, the ratio of the  IR to optical intensity does not 
change much as compared with homogeneously broadened 
transitions, 
but, of course, the threshold conditions do change, so the value of 
injection current is now much larger.

{\bf (ii)}. In this case the population distributions have 
narrow spectral holes burned at the spectral 
positions of optical field modes. This can be relevant for QD lasers 
 or powerful pulsed  multiple QW lasers.  To simplify calculations, 
 let us assume that the absolute values of the Rabi frequencies of 
 two optical fields are equal, $|e_1| = |e_2|$,  all $\g_{ik}$ and 
 $r_i$ are equal to the same value $\g$, and also $u_{21} = u_{31} 
= 
u$.  After somewhat lengthy calculations,  we arrive at a surprisingly simple 
result:  \begin{equation} \label{13} |e| \simeq \frac{\me |e_2| u}{\g 
u_{32}} \left| 0.9 \frac{\w}{\w_1} 
\frac{d^2}{d^2_1}\frac{\kappa_1}{G_1}\frac{G}{\kappa} - 0.1  
\frac{\w}{\w_2} \frac{d^2}{d^2_2}\frac{\kappa_2}{G_2}\frac{G}{\kappa} 
 \right|.  \end{equation} The requirement $|e| < \g$ employed in the 
 derivation of Eq.~(\ref{13}) means that $|e| \ll \me$ in the 
 asymptotics (\ref{11}), since the latter was obtained in the limit 
$\me \gg \g$.

 Expressions (\ref{7})-(\ref{13}) predict the
 rapid growth of the IR intensity,
proportional to the product of optical field intensities. Of course, 
this tendency holds only until the IR intensity reaches the 
saturation value. Above  this value, the IR field begins to deplete 
the electron populations, and its growth becomes  nonlinearly 
saturated. In the optimal case, the maximum  efficiency of 
down-conversion  approaches the limiting quantum value 
corresponding 
to one IR photon per one optical photon. 
For the mid-IR range $5-10$ $\mu$m the maximum IR power is $10-
50$ 
mW if we take the value of $100$ cm$^{-1}$ for IR losses.  
Here 
we assumed the laser with parameter $\eta \sim 0.1$; see the 
discussion after Eq.~(\ref{7}). 
Beyond the restsrahlen region of 
strong phonon dispersion ($\lambda \gtrsim 50$ $\mu$m for 
AlGaAs/GaAs 
structure) the expected IR power is $\lesssim 1$ mW due to 
rapidly 
growing losses. We note, however, that progress in 
fabricating low-loss IR waveguides for semiconductor lasers is 
impressive.  For example, in \cite{SK99}  
waveguide losses 
as low as 14 cm$^{-1}$ at 9.5 $\mu$m wavelength were reported.

Our calculations demonstrate the generation of coherent 
IR emission at intersubband  transitions due to nonlinear wave 
mixing
in standard 
multiple QW or QD laser diodes. The prerequisite for this is 
simultaneous
lasing at two optical wavelengths which provide the necessary drive 
fields.
This mechanism does not require population inversion at the IR 
transition,
and its threshold current is determined by the minimum injection 
current
necessary for the excited-state lasing. 

It is important to note that the proposed parametric scheme seems 
to be viable in the far-IR region ($\lambda \sim 10-100$ $\mu$m) 
using transitions between hole subbands in the valence band. 
Indeed, in conventional lasers the 
increase of IR losses with  wavelength is 
a significant problem 
for generation since it 
makes it difficult to reach 
the lasing threshold. In our case, due to the parametric nature of 
IR generation, the increase in losses will only lead to a decrease in 
output intensity; not a complete loss of IR generation as is the 
case 
in the usual below threshold laser behavior.  For a practical device, 
the goal is to maximize the factor $(\kappa_{1,2}/G_{1,2}) 
 (G/\kappa)$.  Also, in the far-IR spectral range it is relatively 
easy to manipulate the refractive index of the IR mode and provide 
phase 
matching by only a modest doping of waveguide layers.

The authors  gratefully acknowledge 
encouraging and helpful discussions with
D.~Depatie and R. Haden, A. Botero,  Yu. Rostovtsev, and H. Taylor
and thank the Office of Naval Research,
the National Science Foundation, 
the Robert A. Welch Foundation 
and the Texas Advanced Technology Program for their support.



\newpage
\begin{figure}[tbp]
\center{\epsfig{file=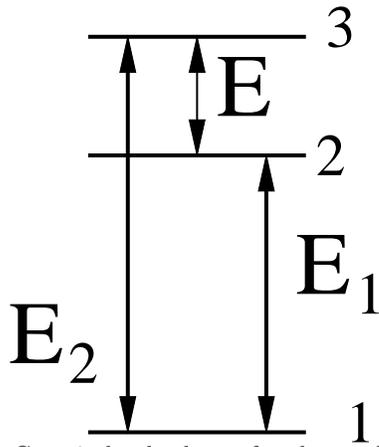, 
width=5cm, angle=0} }
\caption{
Generic level scheme for three-color generation in a field biased 
quantum well: two strong lasing fields $E_1$ and $E_2$ excited at 
adjacent interband transitions $2\rightarrow 1$ and
$3\rightarrow 1$ generate coherent IR radiation $E$ at the beat 
frequency. Wavefunctions in such a skewed quantum well are indicated.
}
\end{figure}

 \end{document}